\documentclass[12pt]{thesis} 

\usepackage[T1]{fontenc}        
\usepackage{textcomp, amssymb}  
\usepackage[latin1]{inputenc}   
\usepackage{setspace}           
\usepackage{anysize}            
\usepackage[all]{xy}            
\usepackage[tight]{subfigure}
\usepackage{multirow}

\marginsize{1.2in}{0.9in}{0.5in}{1.5in} 

\usepackage{ifpdf}              
\ifpdf
  \usepackage{aeguill}          
  \usepackage[pdftex]{graphicx} 
  \usepackage[pdftex]{color}    
  \usepackage[pdftex]{thumbpdf} 
  \usepackage[pdftex,colorlinks,%
              pagebackref=true, 
              linktocpage=true, 
              plainpages=false, 
              bookmarksopen=true,%
              bookmarksnumbered=true,%
              pdfauthor={tewabe chekole },%
              pdftitle={},%
              pdfsubject={Master Thesis},%
              pdfkeywords={Quntum walks ,continuous-time Quntum walks},%
             ]{hyperref}        
\else
  \usepackage[hypertex,
              plainpages=false, 
              linktocpage=true, 
             ]{hyperref}        
  \usepackage[dvips]{color}     
  \usepackage[dvips]{graphicx}  
\fi

\usepackage{makeidx}                       
\makeindex                                 

\usepackage[style=super, cols=3]{glossary} 
\makeglossary                              

\begin{document}

\pagenumbering{roman} 

\ifpdf\pdfbookmark[1]{Title}{label:title}\fi              

\thispagestyle{empty}

\begin{center}

\includegraphics[width=0.15\textwidth]{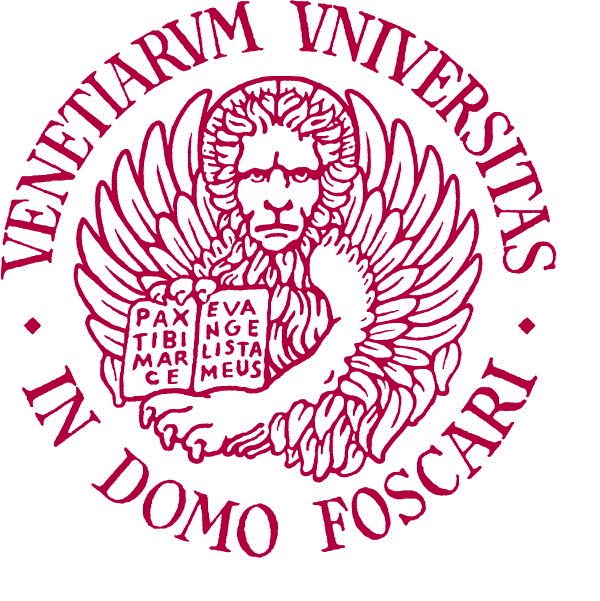}\\[1cm]

\setstretch{1.667}
{\bf\LARGE Ca' Foscari University}

\par Department of Computer Science

\par\vspace{20mm}
{\bf MASTER THESIS}
\par\vspace{2mm}
{\bf\large A  Continuous -  Time Quantum Walk for Attributed Graphs Matching }


\par\vspace{3mm}

\vspace{35mm}
\begin{minipage}{0.4\textwidth}
\begin{flushleft} 
\emph{By:}\\
\emph{TEWABE CHEKOLE 839330}
\end{flushleft}
\end{minipage}
\begin{minipage}{0.4\textwidth}
\begin{flushright} 
\emph{Supervisor: }\hspace{15mm}\color{white}{.}, \\
\emph{\color{black}{Prof.  Andera Torsello}}
\end{flushright}
\end{minipage}

\end{center}


\newpage
\thispagestyle{empty}
\mbox{}

{\doublespacing
\newpage\ifpdf\pdfbookmark[1]{Acknowledgement}{label:ack}\fi\chapter*{Acknowledgments}

I would like to articulate the deepest esteem to my Advisor Prof. Andera Torsello for his relentless support of my professionals study , for his endurance, Motivation and Immense Knowledge. His Guidance assisted me in all the time of writing of this thesis. I couldnt have envisaged having better Advisor and mentor for my professional Study. Next to my advisor, I would like to express gratitude Luca Rossi, he is Ph.D. scholar ,for his support, perceptive comments, and He was always available and He was affirmative and Give bountifully of his time and huge knowledge.He habitually knew where to gaze for the responses to obstacles while premier me the right Source, idea and Perspective for my Question.Also I would like to state articulate deepest admiration to all of my associates in Ca' Foscari University and conjointly All the university groups, i.e. Students, workers, educators, lecturers, for their help and Encouragement  throughout  my stay .

\hspace{0.45cm}Last but not the least; I would like to articulate thankfulness my family: My mother Ethalamu Mengasha , both of my  sisters and  brother  for carrying me Spiritually  throughout  my life.\newpage
\newpage\ifpdf\pdfbookmark[1]{Abstract}{label:abst}\fi     \begin{center}

\vfill
\vspace{5mm}

{\large\bf ABSTRACT}
\end{center}

\hspace{0.45cm} Diverse facets Of the Theory of Quantum Walks on Graph are reviewed Till now .In specific, Quantum network routing ,Kempe \cite{kempe2003quantum},Quantum Walk Search Algorithm, Shenvi ,Kempe and Whaley \cite{shenvi2003quantum} , Element distinctness ,Ambainis \cite{ambainis2007quantum}. associated
to the eigenvalues of Graphs and the use of these relation /connection in the study of Quantum walks is furthermore described. Different  Researchers had contribution  and put their benchmark idea Pertaining with this  research concept.  I furthermore try to investigate recent Application of Quantum walks, In specific  the problem pertained with  Graph matching i.e Matching nodes(vertices) of the Graphs. In this research paper,I consider how Continuous-time quantum walk (CTQW) can be directed to Graph-matching problems. The matching problem is abstracted using  weighted(attributed) Graphs that connects  vertices's of one Graph to other  and  Try to compute  the distance b/n those  Graphs Node's Beside that  finding the matched nodes and  the Cost related to Matching. eventually  measuring the distance between two Graphs which might have different size   then by  using  k-nearest neighbor (k-NN) method try to  classifying those graph based on closest training examples in the feature space.

}

\newpage\ifpdf\pdfbookmark[1]{Table of Contents}{label:toc}\fi \tableofcontents
\newpage\ifpdf\pdfbookmark[1]{List of Figures}{label:lof}\fi     \listoffigures
\newpage\ifpdf\pdfbookmark[1]{List of Tables}{label:lot}\fi       \listoftables

{\doublespacing
\newpage\pagenumbering{arabic}

\newpage\chapter{Introduction}\label{chapter:introduction}

    In  This days computer science communities Give more vigilance and focus for Quantum Algorithm Due to the Features it has and  more affluent representation.  mainly  Because of  they offer substantial  speed-up over classical algorithms and furthermore have more affluent
structure than classical counterparts  .For example, Grover's search \cite{grover1996fast} algorithm is quadratically much quicker furthermore Shors factorization algorithm \cite{shor1994algorithms} is also exponentially faster than renowned classical algorithms. Quantum algorithms furthermore have a more affluent structure than their classical counterparts since they use qubits  rather than bits as the basic representational unit . I n this days Quantum walks have become a common well liked theme of study in different fields  . The quantum computing community is interested in quantum walks due to the role it  plays  in certain algorithms. In quantum computing, quantum walks are the quantum analogue of classical random walks. Like the classical random walk, where the walker's present  state is described by a probability distribution over positions, the walker in a quantum walk is in a superposition of positions. The quantum walk differs from the classical walk in that its state vector is complex-valued rather than real-valued and its evolution is described by unitary rather than stochastic matrices. The evolution of the classical walk is random and a state vector can articulate the probabilities of all the likely outcomes. The evolution of the quantum walk, on the other hand, is deterministic and the randomness only manifests itself when measurements are made. That the state vector of the quantum walk is complex valued allows different routes in the walk to interfere, making amazingly different probability distributions on the vertices's of the graph. Quantum walks have many applications in different areas of aspects. Among that I tried to show one of them and investigate how this well liked study theme is Feasible.

As we understand graph-based or Graphical advances are utilised extensively in computer vision  areas . Much work has been concerned with developing effective graphical representations for diverse submissions and means of performing  this tasks encompass exact and inexact graph equivalent , assessing graph distances , graph embedding  and  graph clustering  broad variety of tasks utilising these representations  . Some of the soonest work which made use of graphical representations of images was that carried out by Barrow and Popplestone[] and Fischler and Elschlager []. Their work demonstrated that high-level object acknowledgement could be achieved by representing images using graphs.
      
        One of the most basic  problem that we need to face  in the graph domain is the of assessing likeness or the distance between the Graphs . How they are alike? Graph matching (GM) performances a central role  in explaining correspondence
problems in computer vision. thus  I tried to investigate this well liked study area  for  this graph matching problems. These studies mostly consider attributed graphs. Graphs are routinely utilised as abstract representations for convoluted scenes, and numerous
computer vision  problems can be formulated as an attributed graph matching  problem,where the nodes of the graphs represent    features of the image and edges correspond to relational aspects  between features (both nodes and edges can be attributed,
i.e. they can encode characteristic/feature vectors). Graph matching/equivalent  then comprises  in finding a correspondence between nodes of the two graphs such that they look most alike when the vertices's are labeled according to such a correspondence.  An intriguing question arises in this context. If we are given two attributed graphs, let say    G and G0, should the optimal match be uniquely determined? In this thesis I used Hungarian algorithm to find the cost  and optimal  between the matching i.e. minimum matching in terms of the cost of matching .The Hungarian algorithm allows a "minimum matching(optimal)" to be found.

In This paper I have to investigate the topic "How Continuous-time quantum walk (CTQW) or Quantum algorithm" can be directed for Graph matching problem .

\section{ Thesis Aim:} The aim of this study is to investigate  one of the application of Quantum walks i.e. Quantum walks for graph matching .I introduce similarity measure between attributed graphs which is based on the evolution of continuous-time quantum walks .In specific, I mainly  focused on the distance b/n the nodes and cost of the matching . Thus, given a pair of graph, i.e. attributed graph, then I conceived  new structure by connecting edges  of one graph to other. On this structure   I defined two continues-time quantum walks which have density operators. Then to define this  similarity assess,  I used  Quantum  Jenson-Shannon Divergence, a assess  which has been presented  as a means to compute the distance between Quantums.then I utilised The Hungarian algorithm to find   the optimal matching. Finally measuring the distance between two Graphs which might  have different size and dimesions  then classifying those Graphs by using  nearest neighbour classification.
\section{ Thesis Overview:}The remainder of this thesis is coordinated as follows : chapter two provides brief overview and essential Introduction about quantum walks continues -time quantum walks, Random walks, Graph theory. In the third chapter i.e. methodology, I introduce the similarity measure and detailed overview how I did this research .In the fourth section  I illustrate the experimental results and in the last chapter deduction is offered..

\chapter{Related Literature}\label{Related Literature}
A quantum walks are the quantum analogue of the classical random walks. In this thesis  I Give , more focus  only for  continuous-time quantum walks. In this chapter I try to  introduce  the concept of a quantum walk, Continuous-time quantum walk (CTQW), Random walks, Graph theory. In Section 2.1, I presented basic notation, concepts and terminology of Graph  especially the idea related to  this thesis. In Section 2.2, I give an overview of random walk and then in section 2.3 I presented basic  definitions and essential introduction to the rudimentary  terminology needed  to understand the suggested  quantum framework like quantum walks and Continuous-time quantum walk (CTQW) and in the last part I recounted, about The Hungarian Algorithm and its significance.

\section{Graph }
Many real-world situations can conveniently be described by means of a diagram consisting of a set of points together with lines joining certain pairs of these points. For example, the points could represent people, with lines joining pairs of friends; or the points might be communication centres, with lines representing communication links. Notice that in such diagrams one is mainly interested in whether or not two given points are joined by a line; the manner in which they are joined is immaterial. A mathematical abstrac- tion of situations of this type gives rise to the concept of a graph.

Recently, several graph based approaches has been proposed to solve problems in computer vision and pattern recognition \cite{tesfayeapplications}, authors in \cite{ AmiShaECCV12, gmmcp, YonEyaPelPraIET2016, YonieyaPAMICDSC_tracker,tesfaye2014multi} formulate multi-target tracking problem in a graph, visual geo-localization \cite{ eyasuPAMIgeoloc, zamir2010accurate, amirshahpami2014} , outlier detections \cite{zemene2016simultaneous}, segmentation \cite{ZemeneAP16,ZemeneAP17,zemene2015path} and so on.

A Graph, G, is defined as an organised pair formed from a set of vertices's, V and a set E of edges. A forming  set of edges is a subset of vertices's, i.e. (x,y)$ \varepsilon $V , where x and y are components  of the set V . To be accurate , if the pair of  A vertex in an edge is unordered, the graph we describe is then known as undirected. All the graph organisations we talk about will be  this kind . An example of an undirected  Graph is shown in fig.1.\newline We define the order of the graph to be the number of  Vertices's, V, and the size to be the number of edges, E. In fig.1.we can see that  some vertices's have more (or less) connections to other vertices's than the other. The degree d of a vertex is defined as the number of edges incident upon it.Graphs offer a convenient way tocomprise diverse kinds of mathematical things.[5]
         Essentially, any graph is made up of two sets, a set of vertices and a set of Edges depending on the specific situation we are endeavouring to represent, although, we may desire to impose limits on the kind of Edge we allow. For some problems we desire the Edge to be directed from one vertex to another; while, in other ones the Edge are undirected\\
A graph G comprises of a finite set V (G) of vertices's and a finite set of E (G) of edges.\\
* Each edge is associated with a set comprising of either one or two vertices's ,\textbf{endpoints}\\
* An edge with just one endpoint is called \textbf{a loop}\\
* Two distinct Edge with the identical set of endpoints are said to be \textbf{parallel}.\\
* Two vertices's that are connected by an edge are called \textbf{adjacent}.\\
* An Edge is said to be incident on each of its \textbf{endpoints}.\\
* Two Edge  occurrence on the identical endpoint are \textbf{adjacent}.\\
* A vertex on which no edges are incident is\textbf{ isolated}.\\
* A graph encompassing no vertices's is called \textbf{empty}; and a graph with vertices's is \textbf{nonempty}.\\

\begin{figure}
\centering
\includegraphics[height=3in,width=5.5in]{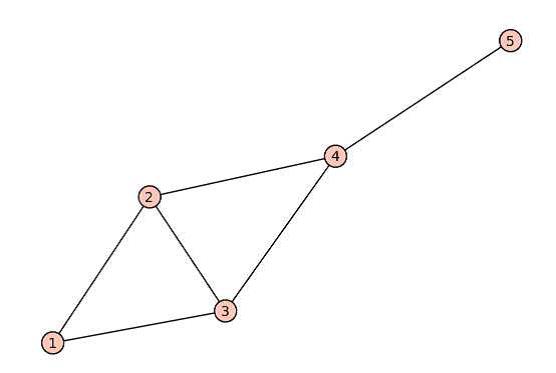}
\caption{An example of a general, undirected graph. It consists of 5 vertices's, of varying degree, and 6 edges}
\label{yourlabel}
\end{figure}


     In this research paper , I only address graphs which have no self-loops. Most will be a single attached  graph, that is there is a path, along edges,
     from any vertex to any other, i.e. there is one connected component. One way to characterise  the connectivity of a structure is by the adjacency matrix. For a graph, G, of  M vertices's, this is a M$\times$ M matrix where the entries of the matrix, A$_{a,b}$, is nonzero
     if an edge exists between vertices's  a and b. In an undirected graph with no self loops, we characterise it\vspace{0.25 cm} \newline \begin{center} A$_{a,b}$= $\bigl\{_{0\ldots\ldots\ldots\ldots otherwise}^{1\ldots.........if\ldots\left\{ a,b\right\} \in E}$\vspace{3mm}
     \end{center}
     
     A similarity  matrix of connectivity is the Laplacian of the graph. This can be defined as: \vspace{3 mm}
     
     \hspace{5cm} L=D-A$_{a,b}$ \vspace{3 mm}\hspace{2.75 cm}                               (2.1)
     
     where D$_{d}$ is the diagonal matrix whose entries Da,a = d$_{a}$, where d$_{a}$ is the degree of
     each vertex. As an example, you can see  both the adjacency and Laplacian matrices of the graph in fig 2.2

\begin{figure}
\centering
\includegraphics[height=4in,width=5 in]{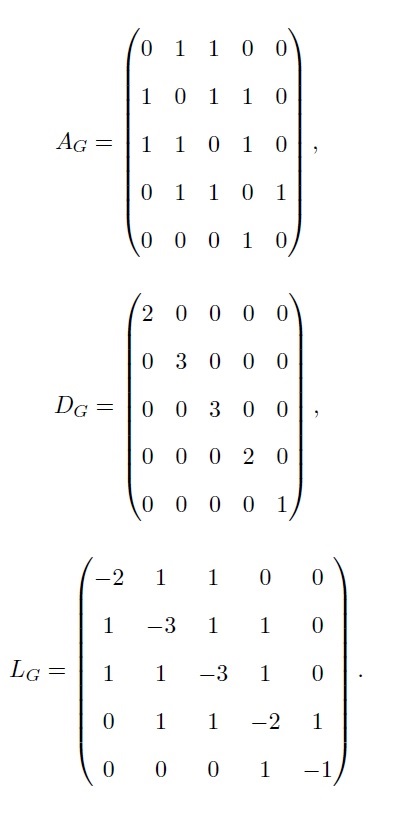}
\caption{the adjacency and Laplacian matrices of the above Graph }
\label{fig:graph}
\end{figure}         

Graphical advances are utilised extensively in computer vision.Much work has been concerned with evolving effective graphical representations for diverse applications and means of accomplishing a wide variety of jobs using these representations. These jobs encompass exact and inexact graph equivalent, calculating graph distances and graph clustering . Some of the soonest work which made use of graphical representations of images illustrated that high-level object recognition could be accomplished by comprising images using graphs.

\section{Random walks }

\hspace{0.45 cm}Random walks are advantageous accoutrement in the assay of the structure  of  a Graphs. The stable state of the random walk  on a graph is granted by the premier eigenvector of the transition probability matrix, and this in turn is related to the Eigen structure of the Laplacian matrix. therefore, the study of random walks  has been the aim of maintained study activity in spectral graph theory. From a functional viewpoint, there have been a number of useful submissions of random walks . One of the most significant of these is the investigation of routing problems in network and circuit theory \cite{torsello2005polynomial} and more latest interest is the use of this notion for random walk by Brin and page \cite{brin2012reprint}to characterise the page Rand  index utilised by the Googlebot search engine \cite{torsello2005polynomial}. In the pattern recognition community, there have been some attempts to use random walks for graph matching . These include the work of Robles-Kelly and Hancock  \cite{robles2004string, robles2005graph} which has used both a standard spectral method  and a more complicated one based on concepts from graph seriation  to alter graphs to strings, so that string matching  procedures may be utilised to compare graphs. Gori, Maggini and Sarti  \cite{gori2005exact} on the other hand, have used ideas borrowed from Page Rank to associate a spectral index with the vertices's of a graph and have then used standard  attributed graph-matching procedures to agree the producing attributed graphs. Random walk methods have a close connection with graph spectral methods for image segmentation . For instance, Melia and Shi \cite{meila2001random} have utilised random walks to reinterpret Shi and Maliks  normalized cut method and continue the procedure to learn image segmentation. Grady  has taken this work one step farther by evolving interactive segmentation schemes founded on the anticipations of random walks. At a higher level, Zhu, Ghahramani and Laerty  \cite{robles2005graph } have performed semi supervised learning utilising random walks on a marked/labeled graph structure. Borgwardt et al. have evolved a  kernel that preserves the path length circulation of a random walk  on a graph, and have utilised this to analyses  protein data . This kernel has been utilised by Neuhaus and Bunke  to kernelise the computation of graph edit distance  and assess the similarity of graphs. eventually, Qiu and Hancock \cite{qiu2007clustering} have shown how the commute times of random walks  can be used to render graph spectral clustering algorithms robust to edge weight errors, and have discovered the submission of the method to image segmentation, multimode motion tracking and graph matching. The commute time permits the vertices of a graph to be embedded in a low-dimensional space, and the geometry of this embedding permits the vertices's to be clustered into disjoint subsets. A random walk consists of two constituents: a states space on which at, any point in time, there is a probability distribution  giving the position of the walk, and a transition function which give the probabilities for transitions between one state and another. As such, random walks are most naturally defined on graphs since the connectivity structure of a graph defnes which transitions are allowed. Additionally, weighted graphs can be utilised if the walk is to be biased in some way. Given a graph and a starting point we select a neighbor of it
at random and move to this  neighbor ;  then we select a neighbor of this point at random and move to it etc. The (random) sequence of points selected this way is a random walk on the graph. A random walk is a finite Markov chain  of links that is time reversible In
fact, there is not much difference between the idea of  random walks on graphs and the idea of finite Markov chains; every Markov chain can be viewed as random walk on a directed graph if we permit  weighted edges. likewise  time reversible Markov chains
can be viewed as random walks on undirected graphs and symmetric Markov chains,as random walks on regular symmetric graphs. Random walks are a model of diffusion which are important in, amongst other areas, statistical physics, applied probability and randomized algorithms .

     Let G=(V, E) be a connected graph with n nodes and m edges. Consider a random walk on G we start at a node v$_{0 }$ if at the t-th step we are at a node v$_{t}$ we move neighbor of v$_{t}$ with probability $\frac{1}{d(v_{t})}$
     Clearly, the sequence of random nodes (v$_{t}$: t=0,1...) is a Markov chain.The node v$_{0 }$  may be fixed but may itself be drawn from some initial  distribution P$_{0}$ . We denote by Pt the distribution of vt \hspace{1 cm}
     \vspace{0.5 cm}  
     
     \hspace{3.5 cm} P$_{t}$(i) = Prob (v$_{t}$= i)\vspace{0.5 cm}\hspace{2.75 cm}                               (2.2)

     We denote by M= (p$_{i,j}$)i,j$\in$V the matrix of transition probabilities
     of this Markov chain. So \vspace{0.5 cm}
     
     \paragraph{\hspace {3.5 cm}p$_{i,j}$= $\bigl\{_{0\ldots\ldots\ldots\ldots otherwise}^{1/d(i)\ldots if\ldots i,j\in E}$}\hspace{3.5 cm} \hspace{0.01 cm}  (2.3)

      \vspace{0.5 cm}Let A$_{G}$ be the adjacency matrix of G and let D denote the diagonal
     matrix with (D)$_{ii}$=1/d(i) then M= DA$_{G}$. If G is d-regular,
     then M=(1/d)A$_{G}$. The rule of the walk can be expressed by the
     simple equation:\vspace{0.5 cm}

            \hspace{3.5 cm}  P$_{t+1}$=M$^{T}$P$_{t} $ \hspace{4 cm} (2.4)
      
     \vspace{0.25 cm} (the distribution of the t-th point is viewed as a vector in R$^{v}$),
     and hence:
     
     \hspace{3.5 cm} P$_{t}$=(M$^{T}$)$^{t}$P$_{0}$\hspace{3 cm} (2.5)\break
     It follows the probability p$_{i,j}$$^{t}$that starting at i we
     reach j in t steps is given by the i,j entry of the matrix M$^{t}$\vspace{1.1cm}
     
\section{Quantum walks }

\hspace{0.45 cm} Quantum walks have been presented as quantum equivalent of random walks,and a good recent review of their properties is given by Kempe \cite{kempe2003quantum}. The behavior of quantum walks is ruled  by unitary ratherthan stochastic matrices. The stochastic
matrix of a classical random walk is such that its columns sum to unity. A unitary matrix, on the other hand, has convoluted applications. For a unitary matrix the squares of the entries in the columns sum to unity. Quantum walks possess a number of interesting
properties not displayed by classical random walks. For example, because of  the evolution of the quantum walk is unitary and therefore reversible, the walks are non-ergodic, and what is more, they do not have a limiting distribution. Quantum walks provide an approach to designing quantum algorithms that lends itself more to physical intuition.There are two different models for the quantum random walk, both of which can be simulated on an arbitrary graph. The first of these is the continuous-time quantum walk proposed by Fahri and Guttmann \cite{farhi1998quantum}. The evolution of the walk is then granted by Schrodinger equation. Fahri and Guttmann \cite{farhi1998quantum} show that if a classical Markov process is able to penetrate a family of trees in polynomial time, then the quantum walk that they define is also able to. They go on to give an example of a family of trees that cannot be penetrated in polynomial time by the classical Markov process but can be by their quantum walk. Unfortunately, they are not able to use this walk to construct a solution to any classically hard problems. By making use of the exponentially faster hitting times that are discerned for continuous-time quantum walks on graphs.An exponential speedup for the hitting time was also furthermore discerned by Kempe \cite{aharonov2001quantum} for the discrete-time quantum walk. The discrete time quantum walk is the quantum analogue of the classical random walk. The quantum type  of the discrete-time quantum walk on an random  graph was formalized by Aharonov, Ambainis and Kempe \cite{aharonov2001quantum}. Kempe considered the walk on the n-dimensional Hypercube and was able to show that the hitting time from one vertex to the one opposite is polynomial in n . In general, the discrete-time quantum walk is not effortlessly  analyzed. Kempe Was able to do so for the hypercube due to its regular structure. She demonstrates how this polynomial hitting time could be of use for a routing problem, and also,  in a later paper, shows that it could be used to solve the 2-SAT problem effieciently  (note that this problem can also be explained  effciently using a classical algorithm).  The many interesting properties exhibited by quantum walks,and the fact that their behavior is dictated by the structure of the graphs supporting them, proposes  that they could be a very helpful  tool for graph analysis. Having said this, there are, as yet,few algorithms for graph matching based on quantum random `walks. One approach is that proposed by Gudkov  \cite{gudkov2002graph}.who suggested modeling the graph as a set of point particles with an attractive force between adjacent vertices's. The system is developed from a set initial state and the set of ordered separation distances calculated. They conjecture that this approach explains the graph isomorphism problem. \vspace{1cm}

        \textbf{2.1.1 A Continuous-time quantum walk (CTQW)}\vspace{0.75cm}{\hspace{1cm}}\\
         \hspace{1cm}Like the classical random walk on a graph, the state-space of the continuous-time quantum
walk is the set of vertices's of the graph. although, the probability of being at a
certain state is given by the square of the amplitude of that state, rather then just the
amplitude of the state (as is the case classically). This permit/allows   destructive as well as constructive
interference to take place. a walk on a given connected graph that is dictated
by a time-varying unitary matrix that relies on the Hamiltonian of the quantum system
and the adjacency matrix. CTQW belongs to what is renowned as Quantum walks, which
also consists of the Discrete-time quantum walk.

       A continuous-time quantum walk (CTQW) on a graph G = (V,E), where V is the set of vertices's (nodes) and E is the set of edges connecting the nodes, is defined as follows:
Let A be the of adjacency matrix of G with elements
\vspace{0.75 cm}

\hspace{3.5 cm} A$_{u,v}$= $\bigl\{_{0\ldots\ldots\ldots\ldots otherwise}^{1\ldots.........if\ldots\left\{ u,v\right\} \in E}$\hspace{2.5cm}                               (2.6)
\vspace{0.5 cm}

D be the degree matrix of G (for which the diagonal entry corresponding to vertex v is degree (v)), and let L = D - A, be the corresponding Laplacian matrix which is postive semi definite. The continuous-time quantum walk on the graph G is then defined by the unitary matrix

\vspace{0.5 cm}
\hspace{3.5 cm}  U(t)=$e^{-itL}$\vspace{0.5 cm}\hspace{4.8 cm}                               (2.7)

Where i is the imaginary unit and t for time.     \vspace{0.5 cm}             

The evolution of the walk is then given by Schrodinger equation, where we take the Hamiltonian of the system to be the graph Laplacian, which yields\vspace{0.5 cm}

\hspace{3.5 cm}  $\frac{d}{dt}$$|\psi_{t}>$ = -iL$|\psi_{t}>$\vspace{0.5 cm}\hspace{3.7  cm}                               (2.8)

Given an initial state $|\psi_{0}>$, we can solve Equation 2.8 to determine the state vector at time t.

\hspace{3.65 cm}$|\psi_{t}>$=e$^{iLt}$$|\psi_{0}>$\vspace{0.5 cm} \hspace{3.5cm}                               (2.9)

Given the Laplacian matrix we can compute its spectral decomposition
L = $\Phi\Lambda\Phi^{T}$ , where $\Phi$ is the n$\times$n matrix
$\Phi$= ($\Phi_{1}$|$\Phi_{2}$|$\ldots|\Phi_{n}$) with the ordered
eigenvectors as columns and $\Lambda$ = diag($\Lambda{}_{1}$,$\Lambda{}_{2}$,$\ldots$$,\Lambda{}_{n}$)
is the n$\times$ n diagonal matrix with the ordered eigenvalues as
elements, such that 0 = $\Lambda$$_{1}$$\leq$$\Lambda_{2}$$\leq$$\ldots\leq\Lambda_{n}$.\break
Using the spectral decomposition of the Graph Laplacian and the fact
that \break
\hspace{0.05 cm}   exp{[}-iLt{]} = $\Phi^{T}$exp{[}-i$\Lambda$t{]} $\Phi$  , we can
finally write 
\vspace{0.65 cm}

\hspace{3.65 cm}$|\psi_{t}>$ = $\Phi^{T}$e$^{-i\Lambda t}$$\Phi$$|\psi_{0}>$ \hspace{2.33 cm}                               (2.10)\vspace{1cm}

             \section {The Hungarian Algorithm }  

                The Hungarian algorithm is an algorithm for explaining a matching difficulty or more generally
an assignment linear programming difficulty. The name was granted by H.W.Kuhn
in acknowledgement of the work of the two mathematicians J.Egervary and D.Konig. The
Hungarian Algorithm is actually a exceptional case of the Primal-Dual Algorithm. It takes
a bipartite graph and makes a maximal matching..\hspace{3 cm}
          \vspace{0.65cm}\newline
          The Hungarian Algorithm:\\
          1. Set S $\leftarrow$ $\emptyset$. (We can furthermore use some other initial matching.)\\
          2. If every vertex in V$_{1}$ or in V$_{2}$ is agreed  in S, then S is a maximum matching and we
          halt.\\
          3. If there are unmatched vertices's in S of V$_{1}$, then go through them in some order constructing
          alternating trees (the procedure of building is not significant as we claimed). If there
          is an augmenting tree, then augmenting the matching S by utilising  the augmenting path we
          have another matching S1. Set S $\leftarrow$ S$_{1}$ and go to 2.\\
          4. If all the alternating trees that have unmatched vertices's in V$_{1}$ as roots are Hungarian, S is
          maximal and we halt.

\chapter{Methodologies}\label{chapter:Methodologies}
   
 Quantum walks are the quantum analogue of classical random walks.  I emphasized only continues-time quantum walks. Given a graph G= (V, E), the state space of the continuous -time quantum walkS defined on G is the set of the vertices's V of the graph. Unlike the classical case, where the evolution of the walker conducted  by stochastic matrix (i.e.  a matrix whose columns sum to unity), in the quantum case the dynamics of the walker is ruled by a complex unitary matrix i.e. the matrix that multiplied by its conjugate transpose yields the identity matrix. therefore the evolution of the walk is reversible which suggests that quantum walks are non-ergodic and do not possess a limiting distribution. G = (V,E), where V is the set of vertices's (nodes) and E is the set of edges connecting the nodes, is defined as follows:
            Let A be the of adjacency matrix of G with elements
            \vspace{0.75 cm}
            
            \hspace{3.5 cm} A$_{u,v}$= $\bigl\{_{0\ldots\ldots\ldots\ldots otherwise}^{1\ldots.........if\ldots\left\{ u,v\right\} \in E}$\hspace{2.5cm}                               (3.1)
            \vspace{0.5 cm}
            
            D be the degree matrix of G (for which the diagonal entry corresponding to vertex v is degree (v)), and let L = D - A, be the corresponding Laplacian matrix . The continuous time quantum walk on the graph G is then defined by the unitary matrix
            
            \vspace{0.5 cm}
            \hspace{3.5 cm}  U(t)=$e^{-itL}$\vspace{0.5 cm}\hspace{4.8 cm}                               (3.2)
            
            Where i is the imaginary unit and t for time.     \vspace{0.5 cm}             
            
            The evolution of the walk is then given by Schrodinger equation, where we take the Hamiltonian of the system to be the graph Laplacian, yields\vspace{0.5 cm}
            
            \hspace{3.5 cm}  $\frac{d}{dt}$$|\psi_{t}>$ = -iL$|\psi_{t}>$\vspace{0.5 cm}\hspace{3.7  cm}                               (3.3)

            Given an initial state $|\psi_{0}>$, we can solve Equation 2.8 to determine the state vector at time t.
            
            \hspace{3.65 cm}$|\psi_{t}>$=e$^{iLt}$$|\psi_{0}>$\vspace{0.5 cm} \hspace{3.5cm}                               (3.4)
            
            Given the Laplacian matrix we can compute its spectral decomposition
            L = $\Phi\Lambda\Phi^{T}$ , where $\Phi$ is the n$\times$n matrix
            $\Phi$= ($\Phi_{1}$|$\Phi_{2}$|$\ldots|\Phi_{n}$) with the ordered
            eigenvectors as columns and $\Lambda$ = diag($\Lambda{}_{1}$,$\Lambda{}_{2}$,$\ldots$$,\Lambda{}_{n}$)
            is the n$\times$ n diagonal matrix with the ordered eigenvalues as
            elements, such that 0 = $\Lambda$$_{1}$$\leq$$\Lambda_{2}$$\leq$$\ldots\leq\Lambda_{n}$.\break
            Using the spectral decomposition of the graph Laplacian and the fact
            that \break
            \hspace{0.05 cm}   exp{[}-iLt{]} = $\Phi^{T}$exp{[}-i$\Lambda$t{]} $\Phi$  , we can
            finally write 
            \vspace{0.65 cm}

            \hspace{4 cm}$|\psi_{t}>$ = $\Phi^{T}$e$^{-i\Lambda t}$$\Phi$$|\psi_{0}>$ \hspace{2.33 cm}                               (3.5)\vspace{1cm} \newline
            
            \textbf{ Quantum Jenson Shannon Divergence:-}A pure state  can be described by the  vector $|\psi_{i}>$.The generalization of probability distributions on density matrices allows to define quantum Jensen Shannon divergence (QJSD). the system is said to be the ensemble of pure states \{$|\psi_{i}>$,P$_{i}$\} . The density  operator(or density matrix) of such a scheme  is defined as\vspace{0.65 cm}

            \hspace{3.65 cm}$\rho$=$ \sum_{i} $ p$_{i}$$|\psi_{i}>$$<\psi_{i}|$\hspace{2.33 cm}                               (3.6)\vspace{1cm}

            The Von Neumann entropy of a density operator$ \rho$ is:\vspace{0.46 cm}
            
              \hspace{3.65 cm}H$_{N}$($\rho$)= -Tr($\rho$$\log$$\rho$)= -$ \sum_{j} $ $\lambda_{j}$ $\log$$\lambda_{j}$\hspace{2.33 cm}                               (3.7)\vspace{1cm}

            where $\lambda_{j}$s are the Eigen values of $\rho$. on the basis of the Von Neumann entropy , I can use  Quantum Jensen Shannon divergence(QJSD) b/n  the density operators $\rho$ and $\sigma$ as
             the Two probability distributions  let say $\rho$ = (p1,$\ldots,$ pn) and $\sigma$ = (q1,$\ldots$$,$qn)on a finite alphabet of size n $\geq$2,Quantum Jensen-Shannon divergence(QJSD) is a measure of divergence between P and Q. It measures the deviation between the Shannon entropy of the mixture ($\rho$ +$\sigma$)/2 and the mixture of the entropies, and is given by:-
            \vspace{0.5 cm}.
            
            \hspace{3.2cm}QJSD($\rho$,$\sigma$) = H$_{N}$$\left(\frac{\rho+\sigma}{2}\right)$-$\frac{1}{2}$H$_{N}$$\left(\rho\right)$-$\frac{1}{2}$H$_{N}$$\left(\sigma\right)$  \hspace{2.5 cm} (3.8) \vspace{1.2cm}
            
            This quantity is always well defined , symmetric and negative definite.It can also be shown that QJSD($\rho$,$\sigma$) is bounded ,i.e\\ \\ 
\vspace{1.2 cm}
            \hspace{4.2cm}
              0$\leq$ QJSD($\rho$,$\sigma$)$\leq$1   \hspace{2.5 cm} (3.9)

\begin{figure}[!ht]
\centering
\includegraphics[height=3in,width=3.5in]{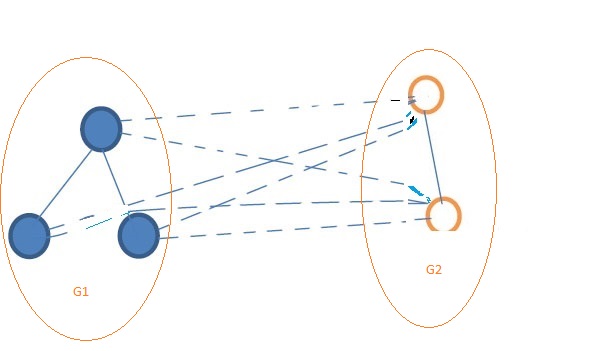}
\caption{Given two graphs G$_{1}$(V$_{1}$,E$_{1}$) and G$_{2}$(V$_{2}$,E$_{2}$), I build  a new graph G=(V,E) where V=V$_{1}$$\cup$V$_{2}$,,E=E$_{1}$$\cup$E$_{2}$$\cup$E$_{12}$ and we add a new edge(u,v) between each pair of nodes u$\epsilon$V$_{1}$ and v$\epsilon$V$_{2}$}\label{fig:graph}
\end{figure}

 \section{ A likeness Measure for Attributed Graphs }  
            Granted two graphs G$_{1}$(V$_{1}$,E$_{1}$) and G$_{2}$(V$_{2}$,E$_{2}$). I construct a new graph G=(V,E)where V=V$_{1}$$\cup$V$_{2}$, E=E$_{1}$$\cup$E$_{2}$$\cup$E$_{12}$ and (u,v)$\epsilon$E$_{12}$ onlyif u$\epsilon$
            V$_{1}$ and v$\epsilon$V$_{2}$. To hand this new structure , I define two continuous-time quantum walks \\
            $|\psi_{t}$$^{-}>$= $\sum_{u\epsilon V}$$\psi_{0u}^{-}|u>$
                                           and $|\psi_{t}$$^{+}>$= $\sum_{u\epsilon V}$$\psi_{0u}^{+}|u>$ on G with starting states.,\vspace{0.75 cm}\\
                                           \hspace{0.5 cm}$|\psi_{0u}^{-}>$=$\left\{ \frac{+\frac{d_{u}}{C}.......if....u\epsilon G_{1}}{-\frac{d_{u}}{C}.......if......u\epsilon G_{2}}\right\} $\hspace{2.5 cm}
                                           $|\psi_{0u}^{+}>$=$\left\{ \frac{+\frac{d_{u}}{C}.......if....u\epsilon G_{1}}{+\frac{d_{u}}{C}.......if......u\epsilon G_{2}}\right\} $\hspace{2.5 cm} (3.10)\vspace{0.75cm}

where d$_{u}$ is the degree of the node u and C is the normalization constant such that the probabilities sum to one .Note that the walk will spread at a speed proportional to the edge weights, which means that given an edge(u,v)$\epsilon$E$_{12}$, the more similar V$_{1}$(u) and V$_{2}$(v)  are , the faster the walker  will propagate along the inter graph connection (u,v).\\
on the other hand , the intra-graph connection weights , which are not dependent on
the nodes similarity , will not affect the propagation speed.               
Given this setting ,  allows the two quantum walks to evolve until a time T and we define the average density operators $\rho$$_{T}$ and $\sigma$$_{T}$ over this time as \vspace{0.75cm}

               \hspace{2.8cm} $\rho$$_{T}$=$\frac{1}{T}$$\intop_{0}^{t}$$|\psi_{t}$$^{-}>$$<\psi_{t}$$^{-}|$dt\hspace{1 cm}   $\sigma$$_{T}$=$\frac{1}{T}$$\intop_{0}^{t}$$|\psi_{t}$$^{+}>$$<\psi_{t}$$^{+}|$dt \hspace{2.5 cm} (3.11)\vspace{1 cm}
              
              In other words, I have  to defined two mixed systems with equal probability of being in any of the pure states defined by the evolution of the quantum walks. then i will prove that, Graph G1 and Graph G2 are isomorphic , the Quantum Jenson-shannon divergnce between $\rho$$_{T}$ and $\sigma$$_{T}$ will be maximum,i.e it will be equal to 1.hence ,we can use this divergence scheme for measuring the similarity between two graphs .

\begin{figure}[!ht]
\centering
\includegraphics[height=3in,width=4.5in]{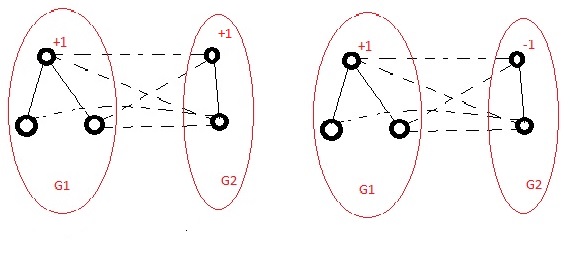}
\caption{Given two graphs which has different Inital point }\label{fig:graph}
\end{figure}

\section {QJSD for attributed Graphs }  
            Given a pair of two attributed graphs G1, G2, then define the quantum Jensen-shannon Divergence between them  as follows:\newline

\hspace{3.7cm}QJSD(G1,G2)=D$_{JS}$($\rho$$_{T}$,$\sigma$$_{T}$) \hspace{2.5 cm} (3.12) \vspace{0.65cm}

 where $\rho$$_{T}$ and $\sigma$$_{T}$ are the density operators defined as in Eq 3.11\\
 \textbf{corollary1} . given a pair of graphs G$_{1}$ and G$_{2}$ then QJSD between those graphs satisfies the following  
               property \\  \vspace{1.2 cm}
            \hspace{4.2cm}0 $\leq$ QJSD(G$_{1}$,G$_{2}$)$\leq$ 1   \hspace{2.5 cm} (3.13) \vspace{0.45cm}

proof. according Eq. 3.12 the QJSD between G$_{1}$ and G$_{2}$ is defined as the quantum Jenson-Shannon divergence between two density operators and the value of quantum Jenson-Shannon divergence is bounded between 0 and 1 .

\section{ QJSD calculation/estimation } 
In this section  I  showed that the solution to Eq 3.12  can be Computed analytically, Particularlly how i did it  computationaly .Recall that $|\psi_{t}>$=e$^{iLt}$$|\psi_{0}>$, then i rewrite Eq. as \\ \vspace{0.45 cm}
\hspace{3.2 cm }$\rho_{T}$=$\frac{1}{T}$$\intop_{0}^{t}$ e$^{-iAt}$$|\psi$$_{0}>$$<\psi$$_{0}|$
e$^{-iAt}$dt \hspace{2.5 cm} (3.14)\\
since e$^{-iAt}$=$\Phi$e$^{-iAt}$$\Phi^{T}$, we can rewrite the previous equation in terms of the spectral decomposition of the adjacency matrix,\\ \vspace{0.45 cm}
\hspace{3.2 cm}$\rho_{T}$=$\frac{1}{T}$$\intop_{0}^{t}$ $\Phi$e$^{-iAt}$$\Phi^{T}$$|\psi$$_{0}>$$<\psi$$_{0}|$
$\Phi$e$^{-iAt}$$\Phi^{T}$ dt \hspace{2.5 cm} (3.15)\\
The(r,c)element of $\rho$$_{T}$ can be computed as\\ \vspace{0.45cm}
\hspace{3.2cm}$\rho_{T}$(r,c)=$\frac{1}{T}$$\intop_{0}^{t}$($\Sigma$$_{k}$$\Sigma$$_{l}$$\phi$$_{rk}$e$^{-i\lambda_{k}t}$$\phi_{lk}$$\psi_{0l}^{-}$)\\ \vspace{0.20 cm}
\hspace{4.5 cm}=($\Sigma$$_{m}$$\Sigma$$_{n}$$\psi_{0m}^{+}$$\phi_{mn}$e$^{-i\lambda_{n}t}$$\phi_{cn}$)dt \hspace{3.5 cm} (3.17)\\

let $\psi$$_{k}$=$\Sigma$$_{l}$$\phi$$_{lk}$$\psi$$_{0l}$  and $\psi$$_{n}$= $\Sigma$$_{m}$$\phi$$_{mn}$$\psi$$_{0n}$$^{+}$ then\\ \vspace{0.4 cm}
\hspace{3.2 cm}$\rho_{T}$(r,c)=$\frac{1}{T}$$\intop_{0}^{t}$($\Sigma$$_{k}$$\phi$$_{rk}$e$^{-i\lambda_{k}t}$$\psi_{k}$$\Sigma$$_{n}$$\phi$$_{cn}$e$^{-i\lambda_{n}t}$$\psi_{n})$ \hspace{3.5 cm} (3.18)\\
which can be finally rewritten as\\
      $\rho_{T}$(r,c)=$\Sigma$$_{k}$$\Sigma$$_{n}$$\phi$$_{rk}$$\phi$$_{cn}$$\psi_{k}$$\psi_{n}$$\frac{1}{T}$$\intop_{0}^{t}$e$^{i(\lambda_{n}-\lambda_{k})t}$dt.\hspace{3.5 cm} (3.19)\\

If we let T $\rightarrow$ $\infty$, Eq.3.19 further simplifies to \\

\hspace{2  cm }$\rho_{T}$(r,c)=$\Sigma$$_{\lambda}$$\Sigma$$_{k\epsilon \lambda}$$\Sigma$$_{n\epsilon \lambda}$$\phi$$_{rk}$$\phi$$_{cn}$$\psi_{k}$$\psi_{n}$

\chapter{ Experimental Results}\label{chapter: Experimental Results}

 \hspace{0.45 cm}In this part, I  evaluate   the performance of the proposed approach and try to  compare it with the value they have. The estimation I utilised for this enquiry
is Quantum Jenson-Shannon Divergence value got by connecting two suggested(
corresponding) node in the given time  .  I did a sequence of experiments   on Different
attributed graph and yields a worth b/n O and 1. for doing this experiment basically  I  used two Graph datasets (i.e COIL AND GRAPHS )  but among  the two dataset I  usually do this computation on the coil dataset , b/c  it has more objects  i.e   100 objects, with 72 views for each object. than the other one , therefore  most of  the tabular output that I putted here is comes from  this data set .\newline

\section{Synthetic Data } 

I start by assessing the suggested approach on set of synthetically developed graphs.
I developed 3 different weighted graph with different size and dimenastion . I took this weighted graph
from coil dataset. In this dataset there is 100 objects with 72 views  for each objects
To this end , I randomly developed 3 different weighted graph prototypes with size
10, 20, 40 respectively .with this synthetic graphs to hand, then try to connect two
Graph lets G and G0 (connect mean just make a connection b/n every vertice of G to every
vertice of G0) this connection yields one graphs G1 as you can seen in fig.4.4.
 
      then after  by utilising  QJSD estimation schema I calculated   QJSD between two Nodes found in different  Graphs and I used Hungarian algorithm  ,to know the cost and assignment of those graphs then I   showed  the  optimal assignment between G and G0 in a tabular way. . fig 4.2 shows the distance (QJSD) b/n the node's of the two graphs with different time interval  . beside this I  randomly pick a graph belonging to one class i.e among the 100 ones and I computed noisy version of that sample Graph. The nosie is applied  to the edges only ,i.e adding or deleting edges, you can see in fig 4.1. then compute QJSD between G and its corrupted versions and  plot it against time , the figure 4.1 shows .computed QJSD between G1 which has 30 nodes  and its corrupted versions with limited time and second one with out limit.(i.e T$\rightarrow\infty$)

\begin{figure}
\includegraphics[height=3in,width=7.5in]{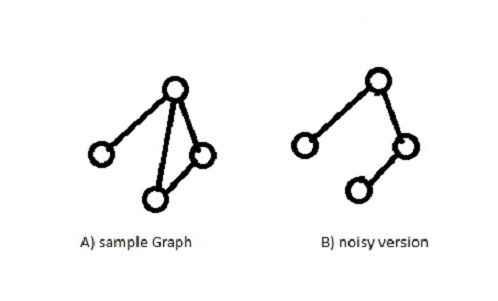}
\caption{Sample Graph and its noisy version }
\label{yourlabel}
\end{figure}

\begin{figure}
\includegraphics[height=3in,width=6.5in]{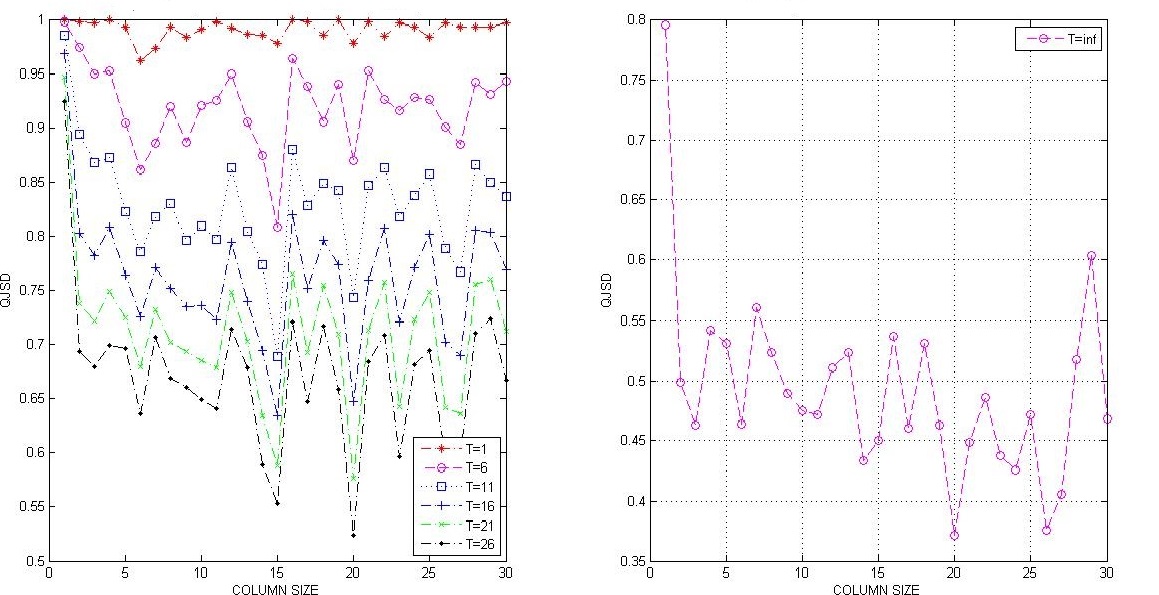}
\caption{QJSD measurement  with different time in a given Graph and its noisy version , the first one  in limited the value of  time and the second one without  }
\label{yourlabel}
\end{figure}

This specific  tabular output  with noise value of  3(three)(i.e adding or  deleting 3 edge) to show the effect at  different points(at different columan) I took the first Row and putted  along with time difference 

\begin{figure}
\includegraphics[height=3in,width=6.5in]{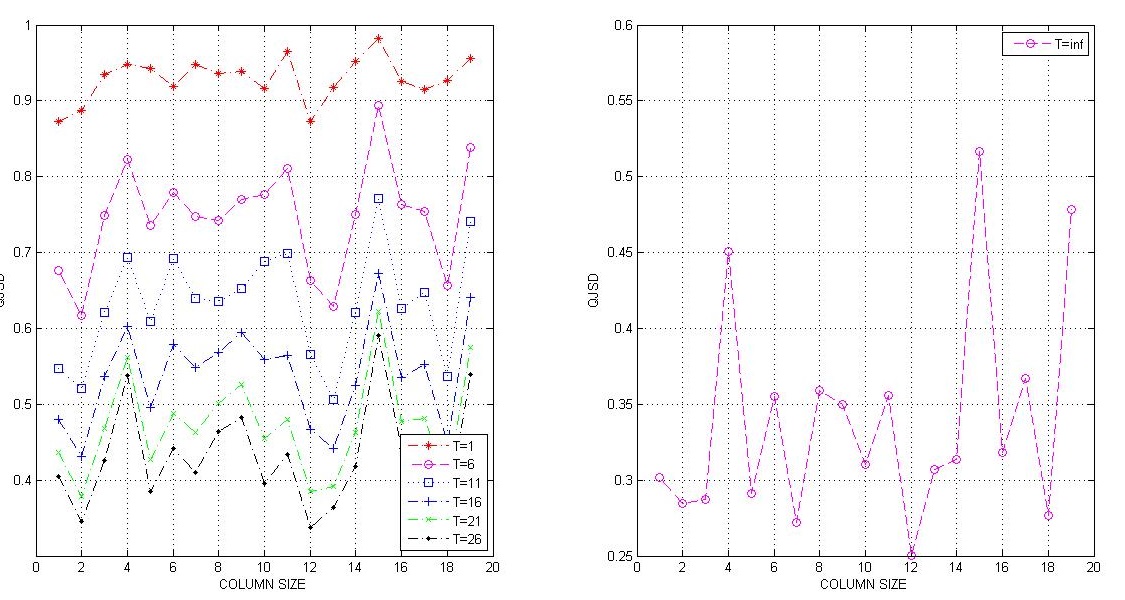}
\caption{QJSD between two nodes found in Different  Graphs with different values of time  }
\label{yourlabel}
\end{figure}

fig 4.2  shows the distance (QJSD) b/n the node's of the two graphs with different time .i.e i selected two Graphs structure  from Coil Dataset. then i connected and making one graphs G . then after  by using QJSD measurement schema I calculated  QJSD between two nodes found in different  graphs and as i did in the  previous one  i took the first row and putted in tabular form along with time difference. 
 
\begin{figure}
\includegraphics[height=3in,width=6.5in]{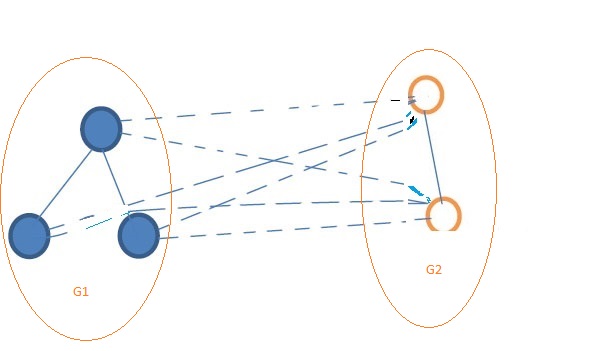}
\caption{Making One Graph by connecting vertices's of two graphs }
\label{yourlabel}
\end{figure}

Fig 4.3   I selected two Graphs structure  from Coil Dataset. then i connected and making one graphs G . then after  by using QJSD measurement schema i calculated   QJSD between two Nodes found in different  Graphs and i used Hungarian algorithm  to know the cost and assignment of those graphs then I  tired  to  show  optimal assignment between G and G0 in a tabular way .

\begin{figure}
\includegraphics[height=3in,width=6.5in]{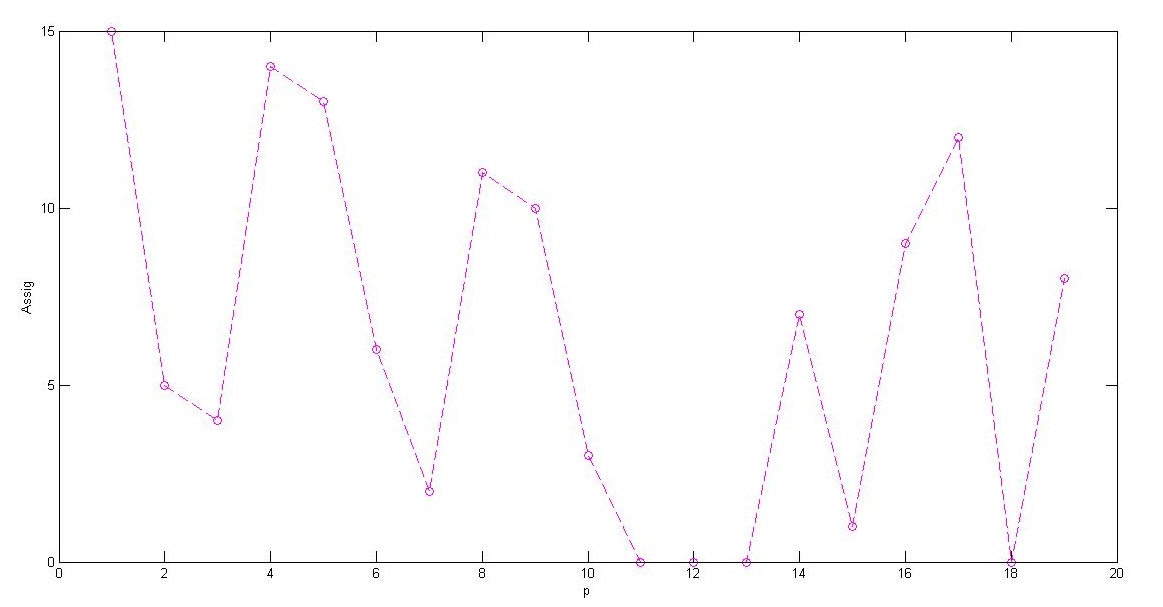}
\caption{Optimal Assignment between Different  Graphs G1 and G2,  }
\label{yourlabel}
\end{figure}


A crucial aspect of matching and recognition problems involves determining a suitable similarity measure between things. In numerous applications  it is needed that such a assess possesses certain properties. In specific it is often desired that a distance assess d fulfills the following properties\\

\hspace{0.025 cm}\textbullet{} d(A, B) $\geq$ 0 (non negativity)\\
\hspace{0.025 cm}\textbullet{} d(A, A) = 0 (identity) \\
\hspace{0.025 cm}\textbullet{} d(A, B) = 0 $\Leftrightarrow$A = B (uniqueness)$^{a}$ \\
\hspace{0.025 cm}\textbullet{} d(A, B) = d(B, A) (symmetry) \\
\hspace{0.025 cm}\textbullet{} d(A, B) + d(B, C) $\geq$d(A, C) (triangle inequality).\\

A distance function satisfying these five  properties is called a metric \cite{torsello2005polynomial}

As second experiment , after knowing  the QJSD Value b/n two nodes of two distinct graphs and b/n one graph and its  noisy version. I should  do some experiment  pertaining with   how this graphs are similar? In alignment to check that ,first i should  calculate  their distance with some formulas then classify those Graphs .The distance between the graph spectra  is computed as pursues  .for each graph G and G0 with adjacency matrix A, what i was doing for this experiment is xoring the two graphs and summing, when i did this may be the graphs has   different dimensions or   different lengths , therefor the vector are all made to be the same length by padding zeros to the end of the shorter  vector.this yields The (i,j)th element of the distance matrix . 
\\ \vspace{0.5 cm}

        \hspace{3.5 cm} d(G$_{1}$,G$_{2}$) = 1 - $\frac{W(G_{12)}}{max(|G_{1}|,|G_{2}|)}$\hspace{3.5 cm} (4.1) \\

       \hspace{3.5 cm}W$_{12}$=$\frac{\Sigma_{_{e1,e2}}xor(G_{1}(e1),G_{2}(e2))}{C(M,2)}$\hspace{3.5 cm} (4.2)\\
 where \\
W$_{12}$= The weight between the Connected Graph \\
G$_{1}$ = Graph1 \\
\hspace{0.25 cm}G$_{2}$ = Graph2\\
M (The Number of Possible Edge b/n the Crossponding Nodes) =  n(n-1)/2  \vspace{0.25 cm}\\ 
 n=Number of Nodes


     After knowing the distance between the two Graphs by using the above
formula I used the k-nearest neighbor (k-NN) method for classifying
those graph based on closest training examples in the feature space.
The nearest-neighbor procedure  is perhaps the simplest of all algorithms
for predicting the class of a test example. The training phase is
trivial , simply store every training example, with its label. To
make a prediction for a test example, first compute its distance
to every training example. Then, keep the k closest training examples,
where k $\geq$ 1 is a fixed integer. Look for the label that is
most common among these examples. This label is the prediction for
this test example.KNN has been involved in much research
because of its simple calculation .
    

A major benefit  of the kNN method is that it
can be used to predict labels of any type.for example  the training and
test examples belong to some set X, while labels belong to some set
Y . The common datasets which have been frequently used in the literature
for evaluating this classification are COIL dataset.this dataset
consists of 100 objects(class) with 72 views for each objects . when
I did this experiment I selected sets as a traning set (50)and test
set (50) randomly from different class of COIL DATASET AND yields 20 percent of the test data
 misclassified  the rest classified  properly.

%
%


\chapter{Conclusion}\label{chapter:conclusion} 
       In this paper ,I have introduce  new way of Graph Matching  on attributed  Graphs Specifcally I tried to investigate  this proposed approch  by utilising the time evolution of Continuous-time quantum walk i.e I conceive new structure by connecting  edges  of one graph to other. On this structure   I define two continues-time quantum walks which have density operators. On this Given pair of Graphs I computed the quantum Jenson-shannon divergence .This yields the quantum Jensen-shannon Divergence  b/n those  Graphs Node's then by using after this I used The Hungarian method  that solves the assignment problem in polynomial time.  then  finding  optimal assignement between those graphs  Beside that  Finding the Cost associated   to Matching. In addition to this ,Ii did some Experiment   pertaining/related to How this graphs are alike?  In order to verfiy  that ,first i should compute  their distance  with the overhead mentioned equations then classify those Graphs by using k-nearest neighbor (k-NN) method . Future work will be aim on revising time parameter more in deepness and furthermore assessing the performance by utilising distinct assessing schema. \\

}

\newpage\phantomsection%
\addcontentsline{toc}{chapter}{\bibname}              
\bibliographystyle{alpha}\bibliography{references}

\newpage\phantomsection%
\addcontentsline{toc}{chapter}{\indexname}                   
\printindex

\newpage\phantomsection%
\addcontentsline{toc}{chapter}{Glossary}                  
\printglossary

\end{document}